\documentclass{article}
\usepackage{spconf,amsmath,graphicx}
\usepackage{parskip}
\usepackage{enumitem}
\setlist{nosep, leftmargin=14pt}
\usepackage[breaklinks]{hyperref}
\usepackage{mwe}

\usepackage{tikz}                  
\usetikzlibrary{positioning, arrows.meta} 

\title{A Novel Vascular Risk Scoring Framework for Quantifying Sex-Specific Cerebral Perfusion from 3D pCASL MRI}    

\name{Sneha Noble$^1$, Neelam Sinha$^1$, Vaanathi Sundaresan$^2$, Thomas Gregor Issac$^1$} 

\begin{document}
\maketitle
\begin{abstract}
The influence of sex and age on cerebral perfusion is recognized, but the specific impacts on regional cerebral blood flow (CBF) and vascular risk remain to be fully characterized. In this study, 3D pseudo-continuous arterial spin labeling (pCASL) MRI was used to identify sex and age related CBF patterns, and a vascular risk score (VRS) was developed based on normative perfusion profiles. Perfusion data from 186 cognitively healthy participants (89 males, 97 females; aged 8 to 92 years), obtained from a publicly available dataset, were analyzed. An extension of the 3D Simple Linear Iterative Clustering (SLIC) supervoxel algorithm was applied to CBF maps to group neighboring voxels with similar intensities into anatomically meaningful regions. Regional CBF features were extracted and used to train a convolutional neural network (CNN) for sex classification and perfusion pattern analysis. Global, age related CBF changes were also assessed. Participant specific VRS was computed by comparing individual CBF profiles to age and sex specific normative data to quantify perfusion deficits. A 95 percent accuracy in sex classification was achieved using the proposed supervoxel based method, and distinct perfusion signatures were identified. Higher CBF was observed in females in medial Brodmann areas 6 and 10, area V5, occipital polar cortex, and insular regions. A global decline in CBF with age was observed in both sexes. Individual perfusion deficits were quantified using VRS, providing a personalized biomarker for early hypoperfusion. Sex and age specific CBF patterns were identified, and a personalized vascular risk biomarker was proposed, contributing to advancements in precision neurology.
\end{abstract}


\section{Introduction}

Arterial Spin Labeling (ASL) is a non-invasive Magnetic Resonance Imaging (MRI) technique designed to quantitatively assess cerebral blood flow (CBF) by magnetically labeling endogenous arterial blood water protons without the need for exogenous contrast agents or ionizing radiation \cite{Lamp5}. The ASL technique involves three key steps: (i) magnetic labeling of arterial blood proximal to the imaging region, (ii) delivery of magnetically tagged blood to brain tissue altering the local MR signal, and (iii) acquisition of paired labeled and control images whose subtraction yields perfusion-weighted maps \cite{Lamp5}. This enables absolute quantification of regional CBF, which serves as a critical biomarker for diagnosing and characterizing neurological disorders such as stroke, Alzheimer’s disease (AD), brain tumors, and epilepsy \cite{Lamp9}. Compared to established methods like positron emission tomography (PET) and single-photon emission computed tomography (SPECT), ASL offers several advantages including non-invasiveness, quantitative measurement capability, repeatability, and absence of ionizing radiation exposure \cite{Lamp8}.

Among ASL variants, 3D pseudo-continuous ASL (pCASL) has emerged as the most widely adopted technique in both clinical and research settings, owing to its optimal balance of sensitivity, safety, and practical implementation \cite{Lamp5}. This approach generates volumetric CBF maps with high spatial resolution, capturing detailed perfusion heterogeneity across the brain \cite{Lamp3,Lamp4}. CBF reflects the delivery of oxygen-rich blood to brain tissue and is tightly coupled to neural metabolic demand and vascular integrity \cite{Lamp27}. Alterations in CBF, whether reductions indicative of ischemia or elevations suggestive of inflammation, are associated with diverse neuropathologies \cite{Lamp28} and provide sensitive markers for early disease detection and progression monitoring.

CBF also serves as a sensitive biomarker of neurovascular function in healthy aging \cite{Lamp26,Lamp29}, where subtle declines often precede measurable cognitive deficits or structural brain changes \cite{Lamp10,Lamp16}. Tracking CBF trajectories across the lifespan is therefore critical for distinguishing normal aging from early neurodegenerative processes, enabling timely preventive interventions.

Sex differences in cerebral perfusion have been consistently documented, with females generally exhibiting higher baseline CBF than males \cite{Lamp1,Lamp7}. These differences are attributed to hormonal influences such as estrogen-mediated vasodilation via endothelial nitric oxide synthase and prostacyclin pathways \cite{Lamp33,Lamp34,Lamp35}, as well as structural brain differences including higher gray-to-white matter ratios in females \cite{Lamp36}. Age-related changes in CBF also exhibit sex-specific patterns; for example, females may demonstrate a more pronounced decline in perfusion in regions critical for cognition, such as the hippocampus \cite{Lamp7,Lamp1}. These sex- and age-dependent perfusion differences are clinically relevant, as they modulate vulnerability and progression of neurological diseases including AD and Parkinson’s disease, where perfusion abnormalities have been shown to differ between sexes \cite{Lamp8,Lamp12}.

While ASL techniques have matured, challenges remain including low signal-to-noise ratio (SNR), sensitivity to motion, and the absence of vessel-selective information \cite{Lamp2,Lamp14}. Diffusion-prepared pCASL (DP-pCASL) has recently been developed to enhance specificity by suppressing intravascular signals, enabling more accurate tissue-level CBF measurement and revealing robust age- and sex-dependent cerebrovascular differences \cite{Lamp13,Lamp1}. Additionally, vessel-encoded pCASL (VEPCASL) provides vessel-specific perfusion maps, improving accuracy in regions with dual arterial supply and offering valuable insights into collateral circulation \cite{Lamp21}.

Standard ASL quantification pipelines such as BASIL and ExploreASL utilize the Buxton general kinetic model \cite{Lamp22} and incorporate motion correction, M0 calibration, and registration to generate voxel-wise CBF maps. However, these approaches may fail to capture finer spatial perfusion patterns and often overestimate CBF in vascular-rich areas \cite{Lamp23,Lamp24}.

To address these limitations, in our study, a 3D Simple Linear Iterative Clustering (3D SLIC) supervoxel algorithm was implemented to segment CBF maps into spatially contiguous and intensity-homogeneous regions. This data-driven method aggregates voxel-level measurements into anatomically meaningful supervoxels, reducing noise and partial volume effects while preserving biologically relevant perfusion heterogeneity. Compared to traditional region-of-interest (ROI) or voxel-wise analyses, supervoxel-based features offer enhanced interpretability and computational efficiency \cite{Lamp25}, which are advantageous for machine learning (ML) \cite{Lamp40} and deep learning (DL) applications \cite{Lamp41,Lamp42,Lamp43,Lamp44}.

The supervoxel-derived features from 3D pCASL CBF maps were leveraged to investigate sex-related differences in cerebral perfusion through a customized DL classifier. The ability of DL models to learn complex, high-dimensional representations enables accurate sex classification based on perfusion patterns, facilitating the identification of sex-specific neurovascular signatures \cite{Lamp17,Lamp19,Lamp20}. Such approaches are critical, as sex is often overlooked as a confounding factor in neuroimaging studies despite its significant influence on brain function, disease susceptibility, and treatment response \cite{Lamp18}. The integration of supervoxel-based regional features with DL thus advances precision neuroimaging by providing interpretable, robust biomarkers with potential clinical utility. Using these approaches, we define a vascular risk scoring framework that will benefit clinicians in predicting vascular diseases in people. 

\section{Methods}

The proposed methodology follows a pipeline to analyze the age- and sex-specific perfusion characteristics of the cohort. 

\subsection{Participants} 

In this study, a publicly available dataset from OpenNeuro comprising 186 cognitively healthy participants was utilized. The cohort includes 89 males and 97 females, with ages ranging from 8 to 92 years. They are of white/Caucasian, Latinx, African American, and Asian origin. 

Each participant underwent DP-pCASL MRI, from which 3D CBF maps were generated. These maps were analyzed to investigate sex classification, under the hypothesis that regional variations in mean perfusion intensity may follow sex-specific trajectories across the lifespan.

\subsection{MR data acquisition}

The 3D CBF images acquired using DP-pCASL MRI were obtained from the OpenNeuro repository. These data were originally collected by Shao et al. (2024) as part of their investigation into the functional integrity of the blood-brain barrier and its variation across age and sex. Imaging was conducted on 3T Siemens Prisma scanners using either 32- or 64-channel head coils \cite{Lamp1}. DP-pCASL acquisition parameters included a spatial resolution of 3.5 × 3.5 × 8 mm\textsuperscript{3}, repetition time (TR) of 4.2 s, echo time (TE) of 36.2 ms, and a field of view (FOV) of 224 mm, with 12 slices and an additional 10\% oversampling. ASL was performed with a labeling duration of 1.5 s. For a post-labeling delay (PLD) of 0.9 s, measurements were acquired at b = 0 and 14 s/mm\textsuperscript{2} across 15 repetitions. For a PLD of 1.8 s, measurements were obtained at b = 0 and 50 s/mm\textsuperscript{2} over 20 repetitions. The total scan time was approximately 10 minutes. T1-weighted structural images were acquired using 3D magnetization-prepared rapid gradient echo (MPRAGE) sequences to enable segmentation and coregistration. MPRAGE parameters included a TR of 1.6 s, inversion time (TI) of 0.95 s, TE of 3 ms, and isotropic spatial resolution of 1 mm\textsuperscript{3}, with a total acquisition time of approximately 6 minutes. Slight variations in these parameters occurred across participating sites.

\subsection{Proposed technique}

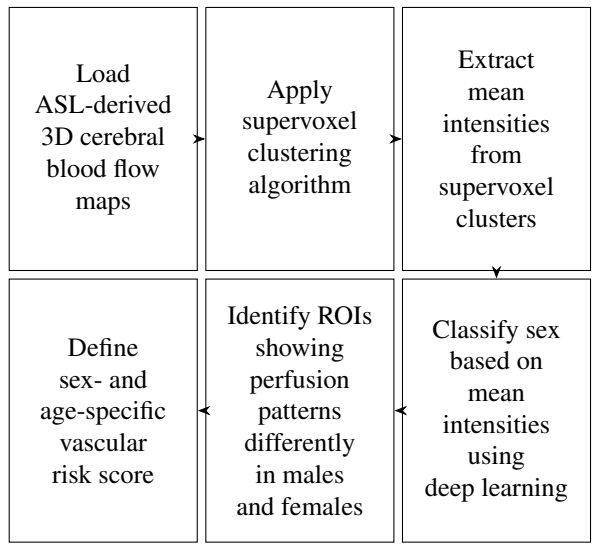
\begin{figure}[ht]
\centering
\begin{tikzpicture}[
    node distance=0.1cm and 0.1cm, 
    block/.style={draw, rectangle, minimum width=2.5cm, minimum height=3.5cm, align=center},
    >={Stealth}
  ]

  \node[block] (A1) {Load\\ASL-derived\\ 3D cerebral\\blood flow\\maps};
  \node[block, right=of A1] (A2) {Apply\\supervoxel\\clustering\\algorithm};
  \node[block, right=of A2] (A3) {Extract\\mean\\intensities\\from\\supervoxel\\clusters};

  \node[block, below=of A3] (B1) {Classify sex\\based on\\mean\\intensities\\using\\ deep learning};
  \node[block, left=of B1] (B2) {Identify ROIs\\showing\\perfusion\\patterns\\differently\\in males\\and females};
  \node[block, left=of B2] (B3) {Define\\sex- and\\age-specific\\vascular\\risk score};
 
  \draw[->] (A1) -- (A2);
  \draw[->] (A2) -- (A3);
  
  \draw[->] (A3) -- (B1);

  \draw[->] (B1) -- (B2);
  \draw[->] (B2) -- (B3);

\end{tikzpicture}
\caption{Proposed framework to obtain sex-based perfusion differences using 3D CBF maps} 
\label{fig:block-diagram}
\end{figure}

\begin{figure}[htbp]
\centering
\includegraphics[width=\linewidth]{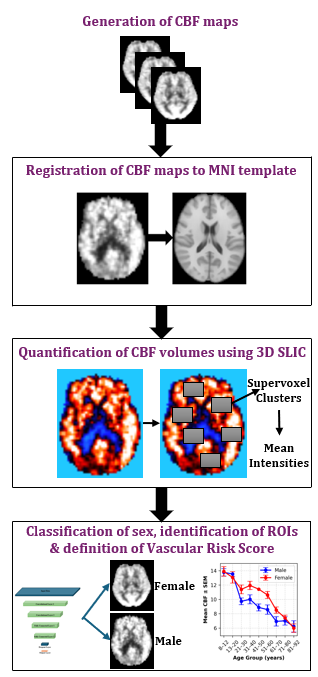}
\caption{Different stages of the proposed workflow: Stage 1—Generation of 3D CBF maps; Stage 2-Registration of CBF maps to MNI space; Stage 3—Quantification of CBF maps using 3D SLIC; Stage 4—Classification of sex, identification of significant ROIs, statistical validation and definition of VRS denoting cerebrovascular risk}
\label{fig:fig2}
\end{figure}

Figure \ref{fig:block-diagram} and Figure \ref{fig:fig2} depict the architecture of the proposed pipeline for classification of participants into males and females using cerebral perfusion imaging. The workflow consists of multiple stages designed to extract and model biologically meaningful features from 3D DP-pCASL CBF maps.

The pipeline begins with data loading and preprocessing, where raw 3D CBF maps are spatially normalized and intensity standardized to ensure inter-participant comparability. Following this, a supervoxel-based clustering technique is applied to segment the brain volume into spatially coherent regions. These supervoxels serve as data-driven regions of interest (ROIs) that preserve anatomical and functional locality while reducing dimensionality.

Within each supervoxel, mean CBF intensity is computed, resulting in a feature vector that characterizes regional cerebral perfusion for each volunteer. These features are then used to train a DL model—specifically, a neural network tailored for classification—designed to distinguish between male and female participants based on perfusion profiles.

After training, the model is evaluated on held-out data to assess classification performance using standard metrics. Finally, the pipeline includes an analysis module that investigates group-level differences in regional perfusion. This includes comparisons between male and female volunteers as well as age-stratified subgroups, allowing for the exploration of how perfusion patterns evolve across the lifespan and vary by sex. Then we analyze the variation in perfusion patterns between men and women as they age, then use the information obtained to develop a vascular risk score (VRS) that determines the vascular risk. 

This comprehensive approach not only aims to achieve accurate classification of sex but also enhances our understanding of the neurobiological differences reflected in CBF patterns.

The key contributions of this study are as follows:

\begin{itemize}
    \item Development of a novel CBF analysis approach - the supervoxel clustering method - that enables simplified and robust analysis of perfusion data;
    \item Design and implementation of a deep learning-based sex classification framework that demonstrates high predictive accuracy using 3D DP-pCASL CBF maps; 
    \item Empirical characterization of sex-related differences in cerebral perfusion patterns across the brain;
    \item Investigation of age-related variations in regional CBF, highlighting the interaction between aging and perfusion dynamics;
    \item Designing a concept called, ``Vascular risk score" to identify the risk of acquiring vascular diseases in males and females. 
\end{itemize}

\subsubsection{Loading of 3D CBF maps}

The analysis begins with the loading of CBF data stored in NIfTI (Neuroimaging Informatics Technology Initiative) format, where each file represents a volumetric brain scan. These 3D CBF maps undergo preprocessing, including intensity normalization, to standardize the data across participants and ensure compatibility with the neural network input requirements. This step is essential for reducing inter-participant variability and enabling the model to learn consistent patterns associated with sex-specific perfusion characteristics. 

\begin{figure}[htbp]
\centerline{\includegraphics{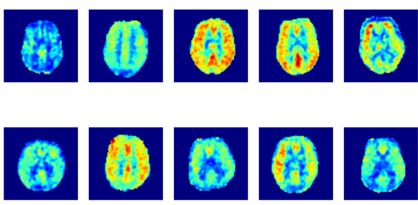}}
\caption{Colored 3D CBF maps where the upper row represents five female participants and the lower row represents five male participants; the bright red colored regions indicate increased perfusion whereas, the dark blue colored regions indicate reduced perfusion, demonstrating variability in regional cerebral perfusion patterns in both sexes}
\label{fig:fig3}
\end{figure}

The 3D CBF maps shown in Figure \ref{fig:fig3} generated from five female and five male participants serve as visual representations of the sex-related variability in regional brain perfusion. The maps showing highly varying perfusion characteristics between male and female participants from the cohort were chosen for representation. These maps reveal pronounced spatial heterogeneities, with specific brain regions exhibiting differential perfusion intensities between males and females. In female participants, certain cortical and subcortical areas display consistently higher CBF values, which may reflect underlying neurovascular or hormonal influences. In contrast, male participants tend to show lower perfusion in corresponding regions, highlighting a pattern that persists across multiple individuals. These inter-individual and inter-sex differences in perfusion patterns are not uniformly distributed but are instead localized to particular regions. 

\subsubsection{Quantification of CBF} 

To ensure anatomical consistency across participants and enable robust group-level analyses, all CBF maps were spatially normalized to the MNI152 standard brain template (Montreal Neurological Institute) using an indirect registration approach via T1-weighted anatomical images. This process was implemented using the FMRIB Software Library (FSL). Linear (affine) registration was employed to align each subject’s CBF map—via the intermediate T2-weighted image—into a common stereotactic space. This alignment enables voxel-wise comparisons of perfusion values across individuals and supports the identification of spatially consistent CBF patterns within anatomically homologous regions \cite{Lamp39}.

Following spatial normalization, quantitative perfusion analysis was conducted through a standardized processing pipeline, incorporating key steps for accurate estimation and summarization of regional CBF characteristics. These steps are detailed in the subsequent section.

\subsubsection*{Supervoxel clustering} 

Supervoxel clustering is employed to segment each 3D CBF map into spatially contiguous and intensity-homogeneous regions, referred to as supervoxels. This technique extends the well-established concept of superpixels from two-dimensional image processing to volumetric neuroimaging data, enabling structured analysis of complex anatomical patterns. In this study, the 3D Simple Linear Iterative Clustering (SLIC) algorithm is utilized, which applies a modified k-means clustering approach in a combined feature space defined by voxel spatial coordinates (x, y, z) and intensity values \cite{Lamp43}. Through iterative refinement, SLIC produces compact and approximately uniform clusters that respect intrinsic anatomical boundaries while preserving perfusion contrast \cite{Lamp15}. 

Supervoxel clustering has emerged as a powerful technique in medical imaging analysis \cite{Lamp41,Lamp42,Lamp43,Lamp44,Lamp45}, offering a structured approach to capture spatial and functional heterogeneity within volumetric data. Its application enhances the accuracy and reliability of perfusion assessment by leveraging voxel similarity and spatial coherence. The key advantages of supervoxel clustering in medical data analysis include:

\begin{itemize}
    \item Enhanced Spatial Coherence: Supervoxel clustering aggregates voxels exhibiting similar signal characteristics, thereby preserving spatial contiguity and enabling the identification of biologically relevant regional perfusion patterns. This grouping minimizes the influence of random spatial noise and enhances interpretability of functional heterogeneity within tissue.
    
    \item Noise Suppression: By averaging voxel-level measurements within each supervoxel, the technique effectively improves the signal-to-noise ratio (SNR). This noise reduction facilitates the detection of subtle perfusion changes that may be critical for early diagnosis and monitoring of pathological conditions.
    
    \item Improved Regional Segmentation: Supervoxel clustering aids in the precise delineation of functional subregions characterized by distinct perfusion properties, such as differentiating ischemic tissue from normally perfused regions. This refined segmentation enhances the accuracy of quantitative assessments and supports targeted therapeutic interventions.
\end{itemize}

Applied to CBF maps generated from 3D pCASL MRI, supervoxel clustering serves to reduce data complexity and suppress voxel-level noise—both common limitations in ASL data—by aggregating voxels with similar perfusion properties into coherent regions. The brain volume of each participant is segmented into 100 supervoxels and the mean CBF intensity is calculated within each supervoxel, resulting in a regionalized perfusion profile. This process is applied uniformly across all participants, generating standardized feature representations for both male and female participants.

These supervoxels are treated as data-driven ROIs, enabling a biologically meaningful and computationally efficient quantification of regional cerebral perfusion. By transforming high-dimensional voxel-wise data into a compact set of interpretable features, this approach facilitates robust inter-participant comparisons and statistical analyses. Importantly, the extracted supervoxel-wise mean intensities serve as input features for downstream ML models, including sex classification tasks. Furthermore, this framework allows for systematic investigation of sex- and age-related variations in perfusion by supporting region-wise group comparisons and exploratory mapping of perfusion heterogeneity across the brain.

\subsubsection*{Extraction of mean intensities of supervoxel clusters}

To capture spatially localized characteristics of cerebral perfusion for meaningful inter-subject and sex-based comparisons, the 3D Simple Linear Iterative Clustering (SLIC) algorithm was applied to partition each normalized 3D CBF map into 100 supervoxels. These supervoxels represent compact, spatially contiguous regions with homogeneous intensity values. This data-driven parcellation provides a flexible and anatomically agnostic alternative to conventional atlas-based region definitions, allowing higher sensitivity to localized perfusion variations.

For each participant, the CBF volume was segmented into supervoxels labeled “Cluster 1” through “Cluster 100.” The mean intensity within each supervoxel was computed, resulting in a 100-dimensional feature vector representing the participant’s regional perfusion profile. The compactness and smoothing parameters for SLIC were empirically set to 10 and 1, respectively, to optimize cross-participant consistency.

All supervoxels, including those with zero or near-zero intensities, were retained to maintain uniform feature dimensionality essential for downstream supervised classification and statistical analyses. This standardized approach facilitates direct comparison of sex-specific perfusion patterns and supports input into deep learning (DL) models.

\subsubsection*{Extraction of mean intensities of the neighboring voxels} 

To capture the spatial context of cerebral perfusion beyond discrete ROIs, we quantified the average CBF intensity within the immediate neighborhoods surrounding the supervoxel clusters identified via SLIC segmentation. While conventional neuroimaging analyses often focus exclusively on ROI-centric metrics, the perfusion characteristics of adjacent tissue provide critical insights into the vascular microenvironment, particularly in studies of cerebral aging and vascular health where interactions between microvasculature and macrovasculature may be pivotal.

For each ROI, defined by its bounding box and centroid coordinates, concentric neighborhoods were delineated by applying increasing margins at radial distances of 0.2 mm, 0.5 mm, 1 mm, and 5 mm from the centroid (assuming isotropic voxel resolution of 1 mm³). A masking procedure was implemented to isolate these peri-regional zones while excluding the core ROI itself, ensuring that measured intensities reflect the perfusion of the immediately adjacent tissue rather than the central cluster. Mean CBF intensities within these surrounding regions were computed individually for all ROIs and participants, yielding spatially resolved perfusion metrics at multiple scales.

Aggregating these neighborhood perfusion values by sex enabled direct comparisons of peri-regional blood flow patterns between male and female participants. This approach is grounded in the hypothesis that sex-specific differences in cerebral microvasculature and macrovasculature—such as capillary density, vessel diameter, and autoregulatory capacity—may manifest not only within focal perfusion hotspots but also in their surrounding vascular territories \cite{Lamp46,Lamp47}. By extending analysis into these perivascular zones, we aim to capture subtle sex-related variations in vascular architecture and function that influence local hemodynamics.

\subsubsection{Classification of participants into male and female groups using a Convolutional Neural Network (CNN)}

In this study, we developed a DL framework to classify the sex of the participant based on regional CBF features extracted from 3D pCASL MRI data. The input to the model consisted of supervoxel-based mean intensity values derived from CBF maps. This data-driven parcellation allowed the model to capture subtle and spatially distributed perfusion patterns potentially associated with sex differences.

The perfusion disparities among men and women, as demonstrated in Figure \ref{fig:fig3}, represent the prominent features leveraged by the CNN to enable robust sex classification. The CNN architecture was specifically tailored to exploit the spatial relationships inherent in the supervoxel feature vectors. Initial convolutional layers were employed to learn localized patterns of perfusion heterogeneity while preserving spatial contiguity across clusters. These convolutional layers were followed by fully connected layers designed to integrate the learned features and perform nonlinear transformations, culminating in a binary classification output corresponding to biological sex.

To evaluate model performance and mitigate bias from dataset imbalances, stratified 5-fold cross-validation was employed, ensuring representative distributions of male and female participants in each fold. During training, metrics such as loss and classification accuracy were systematically tracked to monitor model learning and performance.

\subsubsection{Comparing the results from other classifiers}

To evaluate the performance of different ML approaches on the classification of 3D CBF maps, we implemented several conventional classifiers including Logistic Regression (Linear), Support Vector Machine (SVM) with a Radial Basis Function kernel, Random Forest Classifier (RFC), and XGBoost Classifier (XGBC). These algorithms were selected based on their ability to model both linear and nonlinear relationships \cite{Lamp48}, capture complex feature interactions, and efficiently process structured neuroimaging data.

Each model was trained and evaluated using the same input features and cross-validation strategy as the CNN, allowing for direct comparison of predictive performance across classifiers.

\subsubsection{Age-related variation in cerebral perfusion}

To investigate the influence of age on CBF, we analyzed mean intensity values derived from supervoxel clusters of the 3D CBF maps. These mean supervoxel intensities were computed for each participant and examined as a function of chronological age. The analysis was further stratified by sex to assess potential sex-specific differences in perfusion trajectories across the lifespan.

This approach enabled quantitative characterization of age-dependent changes in perfusion and facilitated identification of sex-related variations in CBF dynamics. Statistical analyses and visualization of age versus mean supervoxel intensity trends were performed to elucidate patterns across different age groups.

\subsubsection{Statistical analysis}

To systematically characterize sex-related differences in cerebral perfusion, we performed statistical analysis on the supervoxel-derived mean intensity features extracted from the 3D CBF maps. Specifically, a one-way Analysis of Variance (ANOVA) was conducted independently for each supervoxel cluster to evaluate whether the mean CBF intensities differed significantly between male and female cohorts. This univariate approach enabled the identification of spatially localized perfusion patterns potentially contributing to sex-specific cerebral hemodynamics.

Prior to conducting ANOVA, assumptions of homogeneity of variance and approximate normality of residuals were assessed to validate the appropriateness of the test. Data transformations or nonparametric alternatives would be considered if these assumptions were violated; however, in the present study, these conditions were met, and thus ANOVA was applied directly.

For each supervoxel, the ANOVA test yielded an F-statistic and corresponding p-value indicating the probability of observing a difference at least as extreme as the measured one under the null hypothesis of no group effect. To address the issue of multiple comparisons, the Bonferroni correction was applied to control the family-wise error rate. Statistical significance was determined using an adjusted p-value threshold of $p < 0.05$.

\subsubsection{Identification of ROIs}

To ascribe biological and anatomical relevance to the statistically significant supervoxel clusters derived from CBF maps, we employed an integrative approach combining data-driven image segmentation with atlas-based anatomical labeling. While the SLIC algorithm partitions the volumetric CBF images into supervoxels—spatially contiguous voxel clusters exhibiting homogeneous perfusion intensities—these clusters lack intrinsic neuroanatomical annotation as they are defined solely by image intensity and spatial proximity.

To contextualize the supervoxel clusters within a standardized neuroanatomical framework, the supervoxel-segmented CBF maps were co-registered to the Brainnetome Atlas—a high-resolution, connectivity-informed parcellation mapped to the MNI152 template. This atlas offers detailed cortical and subcortical delineations, enabling anatomically meaningful interpretation of regional perfusion patterns in relation to established brain structures.

The assignment of supervoxels to anatomical ROIs was operationalized by a frequently-occurring supervoxel-labeling scheme: for each supervoxel, the constituent voxels were overlaid onto the Brainnetome atlas, and the frequency distribution of ROI labels within the cluster volume (obtained from statistical analysis) was computed. The anatomical label most frequently represented within the supervoxel was then attributed as its corresponding ROI. This strategy ensures that each supervoxel is robustly mapped to a singular anatomical region, minimizing ambiguity from partial volume effects or spatial overlap.

This enabled quantification of perfusion metrics within anatomically defined structures. Subsequently, the dataset was stratified by biological sex, allowing for group-wise statistical comparisons using independent two-sample t-tests to identify ROIs exhibiting significant sex-dependent differences in regional perfusion.

\subsubsection{Defining a Vascular Risk Score based on age- and sex-specific CBF intensity}

Age-related reductions in CBF have been extensively documented \cite{Lamp49,Lamp50} and are known to reflect vascular aging and reduced neurovascular efficiency. Moreover, sex-specific physiological differences such as variations in hormonal profiles, vascular tone, and cerebral metabolism, influence baseline CBF levels, with females generally demonstrating higher perfusion than males, especially during their reproductive period. These effects are clearly evident in our analysis (Figure \ref{fig:fig5}), which shows a consistent age-related decline in mean CBF intensity, with sex differences persisting across all age groups.

These findings reinforce the importance of stratifying perfusion measurements by age and sex when evaluating cerebrovascular health. To this end, we introduce a ``Vascular Risk Score (VRS)", a biologically informed, interpretable measure of vascular health derived from age- and sex-specific normative CBF values in a healthy reference cohort.

Rather than relying on z-scoring or percentile cutoffs, we define the VRS based on the empirically observed range of typical CBF variation - specifically, the group mean ($\mu_a^S$) ± one standard deviation ($\sigma_a^S$) for each age group $a$ and sex $S \in \{M, F\}$. This choice reflects an assumption of approximate normality in the healthy population distribution, where roughly 68\% of individuals are expected to fall within one standard deviation of the mean. The standard deviation captures physiological inter-individual variability and thus provides a biologically grounded range of expected perfusion values.

An individual's VRS is determined by comparing their mean CBF to the normative range:

\[
\left[ \mu_a^S - \sigma_a^S,\ \mu_a^S + \sigma_a^S \right]
\]

Values within this interval are considered indicative of healthy or “normal” cerebral perfusion. In contrast, individuals whose mean CBF falls below the lower bound ($\mu_a^S - \sigma_a^S$) are flagged as being ``at risk", with the computed VRS, for cerebrovascular dysfunction. The VRS is given by how far the mean CBF intensities are from the normative.  

This threshold was chosen to reflect a clinically conservative boundary: while some individuals below this range may still be asymptomatic, consistently reduced perfusion has been associated with impaired neurovascular coupling, early small vessel disease, and cognitive vulnerability. 

\subsubsection*{Participant-specific vascular risk score}

To compute the VRS for a participant of a given age and sex, we apply a biologically informed comparison between the individual’s CBF and normative values derived from a healthy reference population stratified by age group $a$ and sex $S$. Let $\text{CBF}_{\text{ind}}$ denote the individual mean CBF (e.g., measured in mL/100g/min) averaged across the brain.

The computation proceeds in the following steps:
\begin{itemize}
    \item Identify the participant’s age group $a$ and sex $S \in \{M, F\}$.
    \item Retrieve the corresponding mean CBF ($\mu_a^S$) and standard deviation ($\sigma_a^S$) from the healthy cohort.
    \item Compare $\text{CBF}_{\text{ind}}$ to the lower bound of the normative interval, defined as $\mu_a^S - \sigma_a^S$.
\end{itemize}

To compare the individual's CBF to the lower bound, the VRS is given by the deviation of the individual's CBF from $\mu_a^S - \sigma_a^S$. Greater this deviation, greater the vascular risk. This can also be used to define a binary classification for vascular status, based on whether the individual's CBF falls below the normative threshold as given below:

\[
\text{VRS} =
\begin{cases}
\text{Normal,} & \text{if } \text{CBF}_{\text{ind}} \geq \mu_a^S - \sigma_a^S \\
\text{At Risk,} & \text{if } \text{CBF}_{\text{ind}} < \mu_a^S - \sigma_a^S
\end{cases}
\]

This threshold provides a conservative criterion for identifying individuals with potentially reduced perfusion, accounting for physiological variability while minimizing false positives.

While the binary classification offers clear clinical interpretability, cerebrovascular function exists on a spectrum. To quantify the extent of deviation below the normative threshold, we define a continuous measure called, ``perfusion deficit" which is given as the following:

\[
\text{Perfusion Deficit} =
\begin{cases}
{\scriptstyle \mu_a^S - \sigma_a^S - \mathrm{CBF}_{ind}}, & {\scriptstyle \mathrm{CBF}_{ind} < \mu_a^S - \sigma_a^S} \\
{\scriptstyle 0}, & {\scriptstyle \text{otherwise}}
\end{cases}
\]

Here, the perfusion deficit reflects how much lower an individual's perfusion is compared to the lower bound of the normative range. A value of zero indicates that there is no detectable deficit, while positive values quantify the magnitude of the potential vascular compromise.

To express this factor on a scale suitable for risk stratification or longitudinal monitoring, we introduce a positive scaling factor $k > 0$, and define the continuous VRS as:

\[
\text{VRS} = k \cdot \text{Perfusion Deficit}
\]

The factor $k$ serves several purposes:
\begin{itemize}
    \item {Unit normalization:} Ensures the score is unitless or mapped to a consistent scale.
    \item {Clinical tuning:} Allows alignment with empirically derived risk categories or intervention thresholds.
    \item {Sensitivity adjustment:} Controls the responsiveness of the VRS to small deviations, which may vary by population or disease context.
\end{itemize}

In practice, $k$ may be chosen empirically based on associations between VRS and downstream outcomes, such as white matter hyperintensity burden, cognitive decline, or vascular biomarkers. This approach provides a continuous, personalized index of perfusion health that enhances sensitivity to subclinical changes and supports individualized risk assessment.

\section{Results}

\subsection{Localized CBF variations revealed by supervoxel features}

To analyze the varying perfusion patterns in men and women and quantify these differences, we analyzed the 3D CBF maps derived from pCASL MRI of healthy participants and the implementation of the 3D SLIC algorithm successfully generated 100 supervoxel clusters per participant. This enabled consistent quantification of localized cerebral perfusion features across the cohort. The extracted mean intensity values within these clusters revealed spatial heterogeneity in CBF distribution, capturing subtle regional differences that may be overlooked by broader anatomical parcellations.

Retaining all supervoxels ensured uniform feature dimensionality, which is critical for subsequent statistical comparison and ML applications. This framework effectively characterized sex-dependent variations in regional perfusion, providing a reproducible and interpretable feature set for investigating neurovascular differences between male and female participants.

Overall, the supervoxel-based quantification demonstrated scalability and robustness in representing fine-grained cerebral perfusion characteristics, underpinning the analytical pipeline for downstream sex classification and physiological variability analyses.

\subsection{Analysis of peri-vascular perfusion variations across sexes}

To better understand the spatial distribution of CBF beyond core regions, we analyzed perfusion intensities in the immediate neighborhoods surrounding the identified ROIs. This allowed us to investigate how cerebral perfusion gradients vary with distance from focal clusters and whether sex-related differences extend into adjacent tissue.

Our analyses revealed a clear monotonic decrease in mean perfusion intensity as the distance from the supervoxel cluster centroid increased. This spatial gradient aligns with physiological expectations, reflecting the transition from high-flow macrovascular inflow regions within the core ROI to the surrounding microvascular beds characterized by lower perfusion levels.

Notably, females exhibited slightly elevated mean neighborhood perfusion intensities compared to males across all radial distances analyzed (0.2 mm, 0.5 mm, 1 mm, and 5 mm). This trend suggests that sex-dependent differences in CBF extend beyond the core supervoxel clusters into adjacent vascular territories.

These findings are consistent with previous studies indicating higher cerebral perfusion in females, particularly in cortical and subcortical areas implicated in cognitive and emotional processing \cite{Lamp1,Lamp3}. The observed spatial heterogeneity in perfusion likely reflects the combined influence of hormonal factors such as estrogen-mediated vasodilation and anatomical differences including higher gray-to-white matter ratios in females \cite{Lamp32,Lamp33,Lamp34}.

The incorporation of neighborhood perfusion metrics at multiple spatial scales enhances the sensitivity of detecting subtle sex-specific vascular differences, potentially serving as early biomarkers for cerebrovascular alterations. Moreover, evaluating perivascular regions may provide crucial insights into the pathophysiological progression of neurovascular and neurodegenerative diseases, where early changes often occur outside of primary lesion sites.

Overall, these results highlight the importance of considering spatially extended perfusion patterns to better capture the complexity of cerebral hemodynamics and sex-related neurovascular variability.

\subsection{Sex classification performance using CNN}

Using the mean intensities of supervoxel clusters derived from 3D CBF maps, the trained CNN achieved a classification accuracy of 95\%, highlighting the discriminative power of localized perfusion patterns in differentiating male and female brains.

Performance metrics obtained through 5-fold stratified cross-validation further support the robustness and generalizability of the model. As shown in Table \ref{tab1}, the CNN consistently demonstrated high precision, recall, and F1 scores across all folds. Precision and recall values were well balanced for both sexes (female precision: 0.94, recall: 0.97; male precision: 0.97, recall: 0.93), indicating minimal classification bias and strong generalization across heterogeneous participant data.

\begin{table}[!ht]
\caption{Prediction performance of CNN model}\label{tab1}%
\centering
\begin{tabular}{|c|c|c|c|}
\hline
& \textbf{Precision} & \textbf{Recall} & \textbf{F1-Score} \\ \hline
Female & 0.94 & 0.97 & 0.95 \\ \hline
Male & 0.97 & 0.93 & 0.95 \\ \hline
Accuracy & & & 0.95 \\ \hline
\end{tabular}
\end{table}

\subsection{Performance comparison across classifiers}

Table \ref{tab2} summarizes the classification accuracies achieved by the alternative models. Ensemble-based approaches, such as the Random Forest Classifier and XGBoost Classifier, showed improved robustness and predictive performance compared to simpler baseline models like Logistic Regression and SVM.

However, the accuracy of all conventional ML classifiers remained lower than that of the CNN-based approach, underscoring the advantage of DL architectures in capturing spatially localized features and learning hierarchical representations from volumetric neuroimaging data.

\begin{table}[!ht]
\caption{Prediction consistency across classifiers}\label{tab2}%
\centering
\begin{tabular}{|c|c|}
\hline
\textbf{Classifier} & \textbf{Accuracy} \\ \hline
Logistic Regression (Linear) & 0.79 \\ \hline
SVM (Radial Basis Function kernel) & 0.87 \\ \hline
Random Forest Ensemble & 0.82 \\ \hline
XGBoost Ensemble & 0.89 \\ \hline
\end{tabular}
\end{table}

These results affirm that classification of participants into male and female groups based on CBF maps derived from pCASL MRI is feasible. Given the non-invasive nature and physiological relevance of ASL imaging, utilizing CBF maps for sex classification offers promising potential for developing personalized neuroimaging biomarkers and advancing understanding of sex-specific cerebral perfusion differences.

\subsection{Age- and sex-dependent variations in CBF} 

\begin{figure}[htbp]
\centerline{\includegraphics{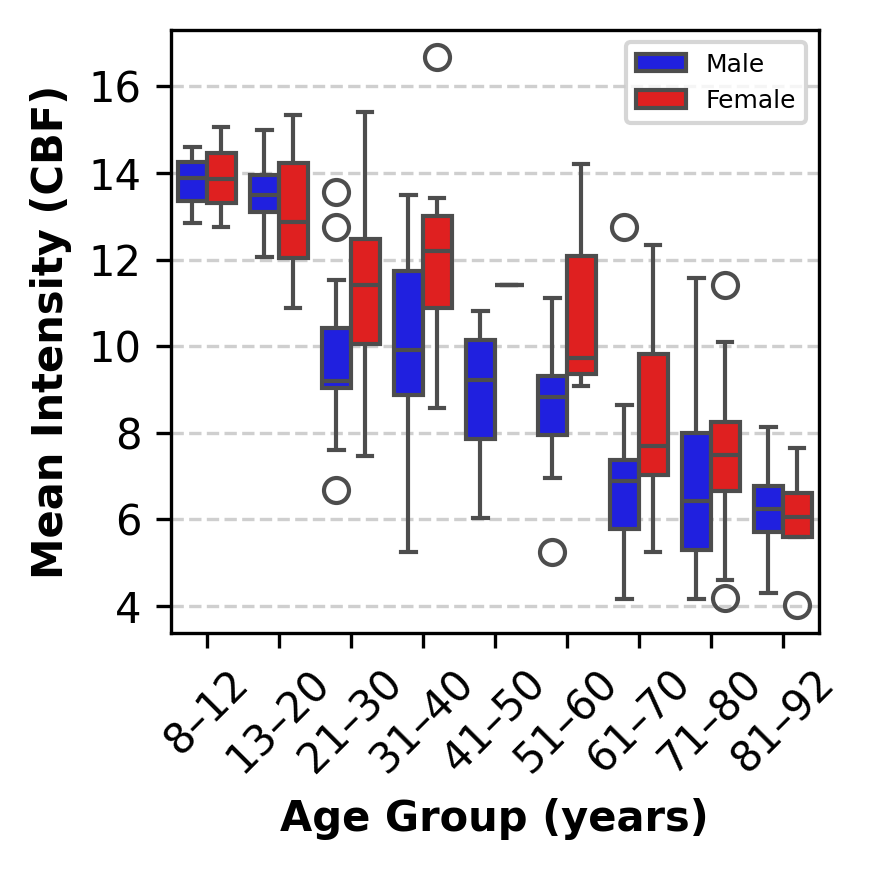}}
\caption{Effect of age on perfusion in males and females}
\label{fig:fig4}
\end{figure}

Consistent with established physiological knowledge \cite{Lamp6}, cerebral perfusion exhibited a characteristic decline with increasing age, reflecting underlying changes in neurovascular coupling, vascular compliance, and cerebral metabolism. Figure \ref{fig:fig4} illustrates the mean CBF intensity values stratified by sex, revealing a gradual reduction in perfusion across the lifespan for both males and females.

Females consistently showed higher average perfusion levels compared to males throughout most age ranges, suggesting relatively preserved CBF. In childhood (8–12 years), perfusion was comparable between sexes. However, beginning in adolescence (13–20 years), females exhibited slightly elevated CBF values, which became more pronounced in early adulthood (21–30 years) and persisted through middle age (31–50 years) and later adulthood (51–70 years). Notably, the sex difference diminished in older adults (71–92 years), where perfusion levels converged.

These results align with previous studies \cite{Lamp1,Lamp6, Lamp30,Lamp31} reporting sex-dependent cerebral perfusion changes, potentially driven by hormonal factors, cardiovascular risk profiles \cite{Lamp38}, and differences in cerebrovascular reserve capacity. The supervoxel-based regional analysis further allowed detection of localized patterns of vulnerability or resilience in brain perfusion.

Overall, these findings underscore the importance of considering both age and sex as critical biological variables in cerebral perfusion research. Understanding these nuanced variations can enhance the development of sex-specific perfusion biomarkers for early detection of cerebrovascular disease and support personalized treatment strategies.

\subsection{Significant sex-dependent perfusion clusters}

The one-way ANOVA identified 17 supervoxel clusters exhibiting statistically significant differences in mean CBF intensities between males and females after Bonferroni correction ($p < 0.05$). These results indicate robust and spatially localized sex-dependent variations in cerebral perfusion.

The identified clusters align with previously reported regions demonstrating sex differences in CBF, highlighting anatomical specificity in sex-related neurovascular regulation. These discriminative features not only corroborate established findings but also provide targeted regions for further neurophysiological and clinical investigation.

Moreover, the statistically significant supervoxel clusters were leveraged as key inputs for subsequent ML classification models, potentially enhancing model interpretability and improving predictive performance in sex classification tasks.

\subsection{Regional perfusion differences between male and female brains} 

After figuring out the significant supervoxel clusters, we identified the major brain regions corresponding to the clusters, that show changes in perfusion in men and women. Table \ref{tab3} summarizes the top six anatomical regions demonstrating statistically significant differences ($p < 0.05$) in mean CBF intensity between male and female participants.

\begin{table}[!ht]
\caption{Top 6 anatomical ROIs that help differentiate females from males based on perfusion changes}\label{tab3}%
\centering
\begin{tabular}{|c|}
\hline
\textbf{ROIs} \\ \hline
Brodmann Area 6 \\ \hline
Brodmann Area 10 \\ \hline
Area V5/MT \\ \hline
Occipital Polar Cortex \\ \hline
Ventral Dysgranular \& Granular Insula \\ \hline
Dorsal Dysgranular Insula \\ \hline
\end{tabular}
\end{table}

\begin{figure}[htbp]
\centerline{\includegraphics{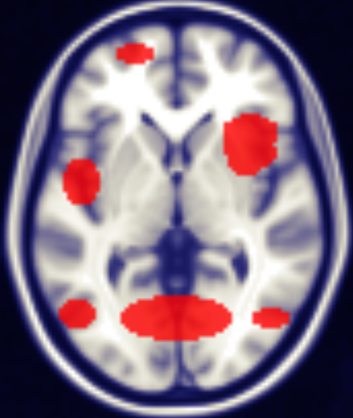}}
\caption{Identified ROIs overlaid on the brain image}
\label{fig:fig5}
\end{figure}

Figure \ref{fig:fig5} displays the identified regions overlaid on the brain image — Medial Area 6, Medial Area 10, Area V5 of the Visual Cortex (also known as the middle temporal area (MT)), Occipital Polar Cortex, Ventral Dysgranular and Granular Insula, and Dorsal Dysgranular Insula. These regions exhibit significant sex-related differences in regional CBF, highlighting their importance as key discriminative features in sex classification models based on perfusion data.

Medial Areas 6 and 10 are part of the Brodmann's areas (BA), a classical cytoarchitectonic mapping of the cerebral cortex widely used to relate brain structure to function. Brodmann Area 6 includes the supplementary motor area (SMA), a region critical for motor planning, coordination, and execution, especially for complex and bilateral movements. Brodmann Area 10, located in the anterior prefrontal cortex, is among the largest and most evolutionarily advanced prefrontal regions, implicated in high-level cognitive functions such as working memory, decision-making, and goal-directed behavior. These medial frontal regions are metabolically demanding and highly sensitive to changes in cerebral perfusion \cite{Lamp52}.

Our findings show pronounced age- and sex-dependent declines in CBF intensity in these Brodmann areas, with males exhibiting a greater reduction than females \cite{Lamp51}. This pattern aligns with established evidence of sex differences in vascular aging and neurovascular coupling, influenced by hormonal factors such as estrogen, which is known to have vasoprotective and metabolic regulatory effects. The progressive CBF decline in these areas likely reflects cumulative cerebrovascular burden that contributes to impairments in motor control and executive functions seen in aging populations. Conversely, the relatively preserved perfusion observed in females may underlie their relative resilience against early neurovascular dysfunction.

Other regions, including Area V5 (MT) and the Occipital Polar Cortex—key nodes in visual motion processing and early visual perception—also show higher perfusion in females, potentially linking to sex differences in visual attention and sensory processing. The Ventral and Dorsal Dysgranular Insula, integral to the salience network and involved in interoceptive awareness, emotion regulation, and cognitive flexibility, exhibit female-biased perfusion increases, supporting sex-specific patterns of affective and adaptive neural processing.

Together, these neurovascular differences emphasize distinct sex- and age-related vulnerabilities and adaptations across motor, cognitive, sensory, and emotional brain networks. The medial Brodmann areas 6 and 10, in particular, emerge as critical biomarkers of cerebrovascular aging and sex-specific neural health, reinforcing their significance in data-driven models for sex classification based on CBF metrics.

\subsection{Normative CBF trends and VRS}

Participant-specific VRS can help determine cerebrovascular risk in an individual. It is an age- and sex-specific metric used to quantify perfusion patterns and deviation of CBF below the normative values. 

\subsubsection{Visualization and Interpretation}

\begin{figure}[htbp]
\centerline{\includegraphics{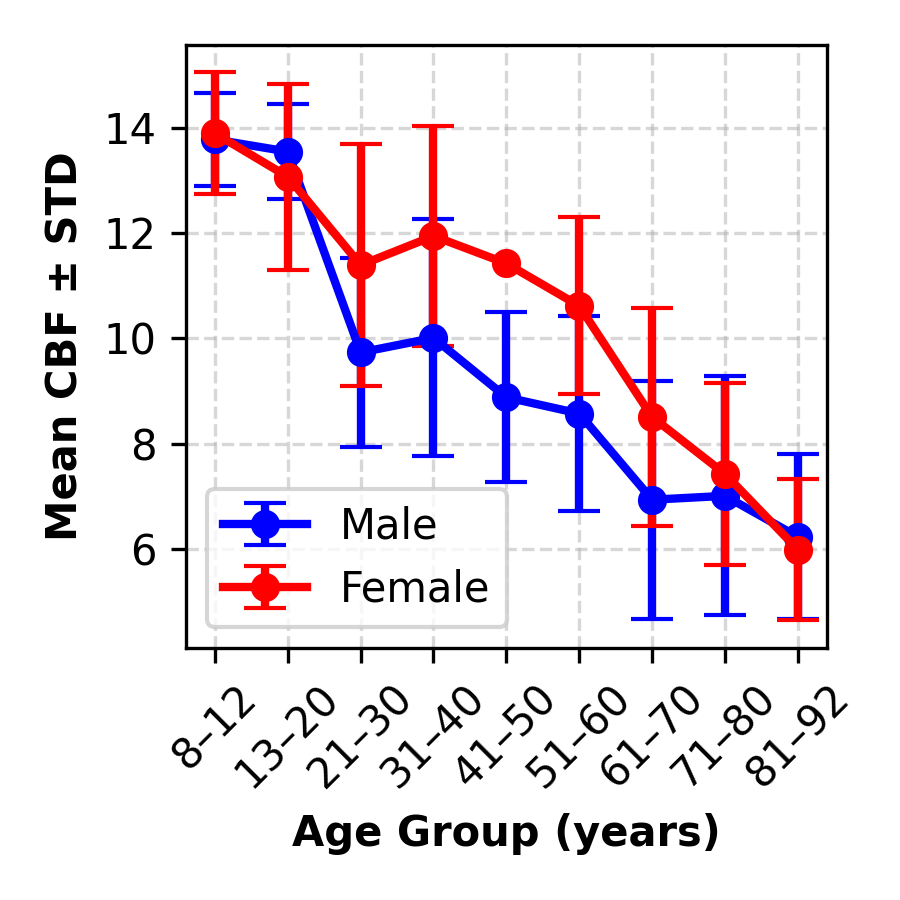}}
\caption{Mean CBF intensity across age groups, stratified by sex. Females (red) and males (blue) are shown separately. Error bars represent one standard deviation from the group mean, defining the normative range used to compute the VRS.}
\label{fig:fig6}
\end{figure}

Figure~\ref{fig:fig6} illustrates the age- and sex-stratified mean CBF intensity values, computed from the healthy cohort. The shaded regions indicate the group-specific standard deviations, defining the normative range against which individual CBF values can be assessed using the proposed Vascular Risk Score (VRS).

Across all age groups, distinct patterns emerged. In early life (ages 8–20), both males and females exhibited elevated and relatively stable CBF values, reflecting preserved neurovascular integrity. Starting in early adulthood (21–40), male participants showed a modest decline in mean perfusion alongside increased inter-subject variability, while females maintained higher and more consistent CBF. This divergence became more pronounced during mid-to-late adulthood (41–80), with males continuing to demonstrate lower CBF values. In the oldest cohort (81–92), both sexes experienced a sharp reduction in CBF, and sex-related differences narrowed considerably—indicating cumulative vascular burden with age.

\subsubsection{Clinical Relevance and Implications}

The VRS derived from this normative dataset was designed to quantify deviations from age- and sex-adjusted mean CBF, enabling the detection of individuals with potentially impaired perfusion. The observed trends support the utility of VRS as a biologically grounded, interpretable, and scalable tool for perfusion-based risk assessment. Specifically, the score offers several advantages for clinical and research applications:

\begin{itemize}
    \item {Biological grounding:} The use of group-level means and standard deviations ensures the score reflects real physiological variation.
    \item {Personalization:} Adjusting for age and sex allows for individualized assessment.
    \item {Interpretability:} The binary classification offers intuitive risk categories, while the continuous version allows finer stratification.
    \item {Clinical utility:} The VRS can be integrated into routine ASL-MRI workflows for early identification of individuals with hypoperfusion.
    \item {Scalability:} The score is easily computable from population data and does not require complex modeling or training datasets.
\end{itemize}

Although not designed as a predictive model for future disease progression, the VRS effectively captures current deviations in cerebral perfusion. It is well suited for longitudinal monitoring, population screening, and integration into multi-modal risk models that combine imaging, genetic, and cognitive data. These findings demonstrate the feasibility and clinical value of the VRS framework in translating population-level perfusion data into actionable, individualized vascular health metrics.

Our analysis demonstrates clear sex- and age-related differences in cerebral perfusion from ASL-MRI data, with females showing higher perfusion than males across most regions and ages. Both sexes experience age-related CBF decline, but males exhibit a steeper decrease and greater variability. Using these normative profiles, we developed a personalized Vascular Risk Score (VRS) to assess perfusion deviations. These results highlight key neurovascular biomarkers of sex-specific brain aging and offer a foundation for individualized vascular health assessment.

\section{Discussion}

Several prior studies have investigated sex-related cerebral differences and sex classification using diverse neuroimaging modalities and computational techniques. Notably, studies have confirmed the presence of sex-specific brain characteristics identifiable through imaging biomarkers \cite{Lamp6,Lamp11,Lamp12}. Extending this line of research, we leverage 3D pCASL–derived CBF maps, combined with supervoxel-based regional intensity features, to develop a robust DL pipeline for sex classification in cognitively normal individuals. In addition, we introduce a novel VRS to quantify individual cerebrovascular risk using perfusion-derived features stratified by age and sex.

The proposed framework employs a supervoxel-based CNN architecture that integrates both micro- and macrovascular perfusion patterns, achieving 95\% classification accuracy. This supports the biological relevance and sensitivity of our method to underlying sex differences in cerebrovascular physiology. Among regions demonstrating significant sex-based perfusion differences, the frontal lobe (BA6, BA10), occipital lobe (Area V5, Occipital polar cortex), and insula emerged as particularly salient. The frontal lobe, involved in executive function, and the occipital lobe, which supports visual processing, exhibited consistently higher perfusion in females. The insula, a key hub for interoception and salience processing, also showed robust sex asymmetries. These findings suggest that sex differences in perfusion are functionally and anatomically grounded and may be shaped by a combination of hormonal, neurovascular, and developmental factors \cite{Lamp6, Lamp33, Lamp34, Lamp35, Lamp36, Lamp37}.

Our results align with those of Shao et al. (2024), who analyzed the same dataset and reported higher CBF in females across age groups, particularly in the frontal and occipital lobes. These reproducible regional differences point toward stable sex-related vascular patterns and reinforce the utility of perfusion-based metrics in identifying sex-specific brain organization. Notably, such disparities have been linked to estrogen-mediated vasodilation, cerebrovascular reactivity, and gray-to-white matter ratios, which differ by sex across development and aging \cite{Lamp32,Lamp33}.

To extract interpretable regional features, we applied the SLIC algorithm to CBF maps. While SLIC-based supervoxel methods have been widely used in various MRI applications—including tissue segmentation in T2-weighted prostate imaging \cite{Lamp41} and brain tumor segmentation in T1-weighted MRI \cite{Lamp42,Lamp43}—their use in ASL data preprocessing remains unexplored. To our knowledge, this is the first study to demonstrate the feasibility of applying SLIC to ASL perfusion imaging. Our results show that SLIC effectively captures spatially coherent regions with similar perfusion profiles, enabling both high classification performance and biomarker extraction from resting-state pCASL data.

The regions identified in our ROI analysis—including medial Brodmann areas 6 and 10, Area V5, the occipital polar cortex, and the dysgranular insula—exhibit significant sex- and age-related differences in CBF. Notably, medial BA6 and BA10, involved in motor and executive functions, show greater perfusion decline in males, consistent with known sex differences \cite{Lamp51,Lamp52} in vascular aging and estrogen’s vasoprotective role. Sensory and affective regions such as V5 and the insula display female-biased perfusion. These patterns suggest that sex-specific CBF variation in these regions may reflect differential neurovascular aging and serve as potential biomarkers of functional decline. 

We also propose a novel vascular risk score (VRS), derived from the mean CBF intensities within supervoxels and benchmarked against age- and sex-specific normative values. This score serves as a non-invasive, individualized index of vascular health. Unlike conventional vascular biomarkers \cite{Lamp49,Lamp50} that require structural lesion detection (e.g., white matter hyperintensities) or advanced acquisitions (e.g., cerebrovascular reactivity testing), the VRS is calculated from standard resting-state ASL scans, enhancing its clinical accessibility. By incorporating demographic stratification, the VRS can detect subtle deviations from normative perfusion patterns—differences that are often obscured by global thresholds or binary classification schemes.

Despite the promising results, several limitations warrant consideration. First, the cohort size (n = 186), although diverse in age, may limit statistical power and generalizability. Additionally, relevant biological covariates—such as hormonal status, cognitive performance, vascular risk profiles, and lifestyle factors—were not included in the analysis, which could confound perfusion estimates. Second, our analysis is restricted to pCASL-derived CBF without integrating structural or functional modalities. Although SLIC supervoxels offer spatially coherent clustering, their anatomical interpretability relies on post hoc atlas mapping, which may introduce inter-subject variability. Lastly, while 5-fold cross-validation was used to mitigate overfitting, external validation on independent datasets remains essential.

Future directions include expanding the cohort to include diverse populations, incorporating additional covariates such as hormone levels, education, and socioeconomic status \cite{Lamp1}, and integrating multimodal data (e.g., fMRI, DTI) to enhance model interpretability and classification robustness. Exploring longitudinal trajectories could help distinguish transient perfusion changes from stable sex-dependent vascular traits. Additionally, domain adaptation and transfer learning strategies may improve cross-scanner and multi-center generalizability. Beyond binary classification, developing age-stratified or multi-class models could reveal nonlinear CBF trajectories across the lifespan and provide deeper insights into developmental and degenerative processes.

From a clinical perspective, targeting perfusion alterations in regions implicated in neurodegenerative disorders such as AD may support early diagnosis and intervention. Combining ASL-derived perfusion maps with functional imaging could further elucidate sex-specific neurovascular responses to pharmacologic or behavioral therapies, ultimately contributing to more precise and personalized treatment strategies. Taken together, our DL pipeline, grounded in interpretable supervoxel-based features, offers a biologically informed framework for precision neuroimaging in both health and disease.

\section{Conclusion}

In this study, we present a framework leveraging 3D pCASL MRI to quantify CBF through supervoxel clustering, enabling anatomically and physiologically meaningful segmentation of brain perfusion patterns. By extracting CBF intensities from spatially contiguous supervoxels in 186 cognitively normal adults, we characterized sex- and age-specific variations in cerebral perfusion. A custom CNN trained on these supervoxel-level features demonstrated robust sex classification performance, achieving high accuracy, thereby, validating the discriminative value of regional CBF metrics as a biomarker. Using the intensities, we developed a novel participant-specific metric called, ``Vascular risk score" that will reveal the risks of acquiring vascular diseases as people age.   

These findings project significant regional differences in cerebral perfusion between males and females, reflecting underlying neurovascular and physiological variations relevant to healthy brain aging. The approach offers a sensitive and non-invasive biomarker framework that captures both microvascular and macrovascular contributions to brain perfusion. Importantly, this methodology has the potential to enhance understanding of sex-specific brain aging trajectories and facilitate early detection of neurodegenerative conditions such as AD.

\section{Conflict of Interest Statement}
All authors declare no conflicts of interest.

\section{Data Availability Statement}
The 3D pCASL MRI data used for the study is available in the open-source database, OpenNeuro. The source code of the proposed method can be made available on request from the authors. 

\section{Acknowledgements}
This study was conducted as part of the Funding for Aging Brain Research in Collaboration (FABRIC) Grant, initiated by the Centre for Brain Research, Indian Institute of Science, Bengaluru (authors S.N., N.S., and T.I.), and the Indian Institute of Science, Bengaluru (author V.S.). Author V.S. is supported by the DBT/Wellcome Trust India Alliance Fellowship [IA/E/22/1/506763] and the Council of Scientific \& Industrial Research (CSIR), India, under its ASPIRE (Women Scientist Scheme) program [25WS(013)/2023-24/EMR-II/ASPIRE]. The authors thank the Director and Administration of the Centre for Brain Research, Indian Institute of Science, Bengaluru, for their support throughout the study.

\section{Supporting Information}

\begin{table}[!ht]
\caption{Supplementary Details}\label{tab4}%
\centering
\begin{tabular}{|c|p{4cm}|} 
\hline
\textbf{File name} & \textbf{Description} \\ \hline
Supplementary\_Information.docx & 
Figure S1: This illustrates the distribution of participants across the defined age groups, providing a visual representation of the demographic composition of the study sample. \newline 
Table T1: It represents the summary of the CNN-based DL model we have applied on the 3D CBF maps. \\ \hline
\end{tabular}
\end{table}

\bibliographystyle{IEEEbib}
\bibliography{strings,refs}

\end{document}